# PRODUCT LINE DEVELOPMENT ARCHITECTURAL MODEL

Ankit Chaudhary, Basant K.Verma
Dept. of Computer Science
Birla Institute of Technology & Science, Pilani
Rajasthan, India
{ankitc.bitspilani, bits.basant}@gmail.com

Jagdish L. Raheja
Digital Signal Processing Group
CEERI Pilani
Rajasthan, India
jagdish@ceeri.ernet.in

*Abstract—* Products with new features need to be introduced on the market in a rapid pace and organizations need to speed up their development process. The ordinary way to develop products, one at a time, is not time efficient enough and is costly. Reuse has been suggested as a solution, but to achieve effective reuse within an organization a planned and proactive effort must be used. Product lines are the most promising technique and it increases productivity and software quality and decreases time-to-market. This paper describes the architecture of product line engineering process and also addresses what the design issues of product line architecture are and how a UML profile looks like for a product line by referring to the basic aspects of a case study, CelsiusTech in its Naval Product Line, SS2000.

## 1. Introduction

Software engineers have designed software systems as one system at a time since the beginning, and each software product involves investments in requirements analysis, architecture and design, documentation etc. More and more organizations realize that they cannot afford to develop multiple software products as one product at a time. They are also pressured to introduce new products and add functionality to existing products at a rapid pace to be able to compete at the market. These goals are hard to meet when designing one system at a time.

Most organizations today usually derive new systems from previous instances to speed up the process. But to reuse similarities between systems in the most efficient way a product line approach might be the right answer to an organization. The approach uses a common set of core assets to modify, assemble, instantiate, or generate multiple products and is referred to as a product line. Such a product line approach involves building a product line as a product family. On the similar lines, a Software Product Line [2, 6] can be viewed as a collection of products that are similar in some important respect yet systematically different in others (for example, successive revisions of a single application, versions of an application for different host platforms, versions with varying features). To speak technically, a *Software Product Line* is a set of software-intensive systems sharing a common, managed set of features that satisfy the specific needs of a particular market segment or mission and that are developed from a common set of core assets in a prescribed way.

Through out the paper, we discussed the product line engineering process and various activities involved in the process, design issues of product line architecture and finally UML profile for product line has been mentioned in brief.

### 1.1 Related Work

The ideas and concepts mentioned in this paper follow on from work on product lines by Linda M. Northrop of Software Engineering Institute [3], Caroline Nyholm [1], the paper by Tewfik Ziadi on UML profile for product lines [8] and the technical report on product line development by Michael Krebs [5]. The case study discussed in this paper is referred from the technical report on a case study in successful product line development by Lisa Brownsword and Paul Clements [4].

## 2. Basic Terminology of Product Line

Each system in the product line is a product in its own right. However, it is created by taking applicable components from a common asset base, tailoring them through preplanned variation mechanisms, adding new components as necessary, and assembling the collection according to the rules of a common, product-line-wide architecture. Every software product line has a predefined guide or plan that specifies the exact product building approach.

Different set of terms are used to convey essentially the same meaning of product line. Some practitioners might refer to a product line as a *Product Family*, to the core asset set as a *Platform*, or to the products of the software product line as *Customizations* instead of products. Others use the terms *Domain* and *Product Line* interchangeably, which can be distinguished. A domain is a specialized body of knowledge, an area of expertise, or a collection of related functionality. Core asset development is often referred to as *Domain Engineering*, and product development as *Application Engineering*.

Regardless of terminology, software product line practice involves strategic, large grained reuse, which means that software product lines are as much about business practices as they are about technical practices. Using a common set of assets to build products requires planning, investment, and strategic thinking that look beyond a single product.

Reuse, as a strategy for decreasing development costs and improving quality, is not a new idea. However, past reuse agendas, which focused on reusing relatively small pieces of code or opportunistically cloning code designed for one system for use in another, have not been profitable. In a software product line approach, reuse is planned, enabled, and enforced. The reusable asset base includes artifacts in software development that are costly to develop from scratch.

Numerous organizations in various industries have reaped significant benefits using a software product line approach for their systems. To mention, at the highest level of generality are three essential and highly iterative activities [3] that blend technology and business practices. Fielding a product line involves *core asset development* and *product development* using the core assets under the aegis of technical and organizational *management.*

## 3. Product Line Engineering Process

As mentioned above, core assets of product line have to deal with the differences between these applications, which mean that possible changes must already be anticipated when developing assets. These differences are so-called variation points. During development these variation points must be foreseen and a proper mechanism to enable variability must be instantiated, so that new applications can easily be derived from the core assets. Because this procedure has totally new demands to the engineering practices, product line development also defines its own software engineering process to support the product line specific peculiarities.

Software product line development [5] consists of several different activities as can be seen in figure 1. The starting point of product line development is Scoping. Scoping determines the set of products and features which can be built within the product line. Information about existing products which in the future should be covered by the product line may be useful for Scoping. At the end of the Scoping activity there is also a decision whether a product line should be developed at all. The reasons for not starting a product line can be that, for instance, the product line would comprise too few applications or that the product line is economically not acceptable.

To see how scope can be determined, we considered the case of CelsiusTech, a leading supplier of command-and-control systems within Sweden's' largest, and one of Europe's leading, defense industry groups. In this case, scope is determined based on the integrated system that unifies all weapons, command-and-control, and communication systems on a warship with a set of features like fire control, warfare systems etc and various products like corvettes, vessels, frigates etc. This Product Line corresponding to CelsiusTech Navy applications is named as Ship System2000 (SS2000) [4]. The next activity in product line development is Domain engineering respectively Core Asset Development. This comprises the sub activities of Domain Analysis, Domain Design, and Domain Implementation. In the analysis phase the requirements for the product line as well as the necessary assets themselves are determined. Information about existing products can be useful for finding the requirements and the variation points of the product line. In the domain design phase the product line structure is defined. This will also be the structure of all applications within the product line scope. Finally the core assets are implemented. The core assets now form the product line infrastructure which is used for developing new applications/variant.

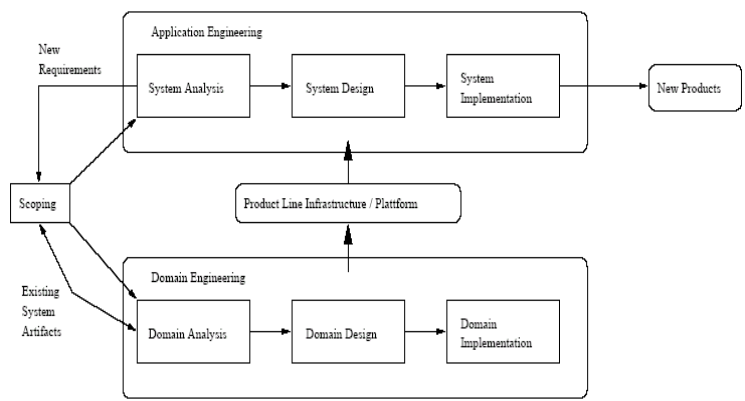

Figure 1.  Overview of the Product Line Engineering Process

In SS2000[4] product line of CelsiusTech, the requirements and variations is discovered by tracking and analyzing the existing products' features like of coastal corvettes', multi role patrol vessels', frigates', submarines' etc., of various countries like Sweden, Australia, and Republic of Oman etc. The most important of these requirements, those vary from submarines to vessels are generally performance, modifiability, safety, reliability, availability and testability. During the design phase of SS2000[4], they designed the operating environment and physical architecture with different processors like Gun Processor, Radar Detector processor, communication processor etc for different purposes connected over a Dual Ethernet LAN.

After setting up the infrastructure new products can be derived during the Application Engineering activity. Decisions have to be made at the variation points to definitely define the requirements. In the system design phase the architecture of the product line will be adapted to the new application's needs while in the last phase the new application will be instantiated according to the decisions made for the variation points.

With the physical architecture of SS2000 [4], the new products are derived. Responding to the continuously arriving inputs and controlling the weapons under tight deadlines is one of the most important requirements that are to be met for such products. Generally, systems built from the product line vary greatly in size, function and armaments and SS2000 [4] is no exception. The basic architecture of SS2000 [4] is adapted to such changes based on the system's needs and variation points such as interface units, underlying processors specific to the application, operating systems that are compatible to the whole system etc.

It has to be noted that the whole process is iterative. This means that every new product has new demands for the product line assets. If a new feature or variation will be identified during product instantiation, it has to be decided. This leads to an evolution of the scope and the product line platform during the whole life cycle.

## 4. Design of Product Line Architecture

Product-line architecture [1] is a software architecture that will satisfy the needs of the product line in general and the individual products in particular within the scope. The product-line architecture specifies the structure of the products in the product line.

A product line can be designed from the ground up or can be built using assets from previous efforts. A designed product line will likely be better for meeting long-term goals, but a mined product line may provide a shorter time to market if the at-hand components are rich enough.

There are differences between the design of architecture for a product line and architecture for an individual product. One difference is that product-specific features need to be considered when designing the product-line architecture. Design decisions made for the product-line architecture can make it impossible for specific products to implement their features if the product-specific features are not considered. This concerns both quality attributes and functional requirements. There are different methods of architectural design: Architecture Based Design (ABD) and Functionality based architectural design [1]. These methods are discussed in the subsections.

### 4.1 Architecture Based Design

Architecture Based Design is a method for designing the high-level software architecture or a product line. It is difficult to design an architecture for a product line because detailed requirements are not known in advance. Since there also are variations between products, the ABD method fulfills functional, quality, and business requirements at a level of abstraction to allow the variations.

The initial stages of architecture design are where the most fundamental design decisions are made. If they are wrong in some way it will be hard to correct them later. To prevent this the architect needs a disciplined design method that provides a strategy for handling the uncertainty in requirements, provides guidance in organizing the decisions made during the design process and make clear why the steps of the method exist and how they relate to each other. The ABD method makes it possible to start with the design activities as soon as the architectural drivers have been determined. This can speed up the process, since the determination of requirements and analysis activities does not have to be complete and can be performed in parallel with the design activities.

There are three *foundations* in the ABD model. The first one is decomposition of functions where well-established techniques based on coupling and cohesion is used. The second foundation is the realization of quality and business requirements through the choice of architectural style. The third and last foundation is the use of software templates, which is a new concept for design methods but has been utilized in the construction of some systems.

The architectural solution of SS2000, naval product line of CelsiusTech [4] mentioned the decomposition of various functions of CelsiusTech by naming the modules as system functions and system function groups. A *system function* is a collection of software that implements a logically connected set of requirements. It is composed of a number of ADA code units. A *system function group* comprises a set of system functions and forms the basic work assignment for the development team of CelsiusTech. SS2000 consists of about 30 system function groups, each comprising up to 20 or so system functions. They are clustered around major functional areas like command, control and communications, weapons control and Human Computer Interface. The quality and business requirements are looked after by the architectural styles as mentioned in detail if we refer to [4].

There are three different *views* in the ABD model, used for different things. The logical view records the responsibilities and conceptual interfaces for the design elements to see their role within the system. The concurrency view is used to examine the system when parallel activities are performed, like multiple users and start-up. The last view is the deployment view, which represents nodes in a computer network (the physical structure of the system). This view is only used for systems that execute on multiple processors.

The architectural solution of SS2000 [4] discussed various views to describe the architectural based design of CelsiusTech' naval product line namely process view, layered view and module decomposition view. The responsibilities and roles of various design elements are captured in the layered view by grouping the modules based on the type of information they encapsulate. The layers are ordered with hardware dependent layers like information regarding operating system, LAN, Inter process communication, base system hardware etc., at one end and application specific layers like target tracking, fire control, ships information, database etc at the other end.

We know that the smallest unit of implementation is ADA program and each ADA program runs at most one processor. A program may consist of several ADA tasks. Systems in SS2000 product line can consist of up to 300 ADA programs. Here comes the process view into picture. These ADA task facilities are used to implement the threading model and Inter process communication plays an important role for data transport between ADA applications. Having a process view at all means that the performance tactic "introduce concurrency" has been applied.

**4.2 Functional architectural design**

Functional architectural design is concerned with the definition of the product context, the identification of archetypes and the description of product instantiations. It differs from the ABD model in that the requirements must be complete before beginning.

Since both functional and quality requirements were defined and scoped in the earlier steps, and the method assumes that the quality requirements are ignored in the start, a slight modification of the model is necessary. According to the model the first step is to define a requirement specification for the product-line architecture that combines the functional requirements for each feature in one set of functional requirements. The modification is to perform the same activity for the quality requirements. And finally the features are reorganized into a set of functional and a set of quality requirements is performed for each product, since we have to evaluate the product-line architecture with respect to its suitability for the products in the product line scope.

The next thing to do is to define the product context. This is difficult since the products in the product line may be very diverse. The contexts in which products in the product line operate are not necessarily specified for the product line as a whole. Some product context aspects must be addressed by the product-line architecture to avoid product-wide effects while other aspects have minimal effects on the architecture and can be handled for just that product. It is up to the software architect to decide which approach to choose.

Now the identification and definition of archetypes are to be performed. The archetypes represent the core concepts used for modeling the software architecture and for describing the product instantiations. They are also used to represent the commonality between the products in the product line. To be able to identify the archetypes optimal for the product-line architecture they have to be based on the product-line requirements and at least on the primary product specific requirements as well and if there are many product-specific requirements it may be necessary to identify product-specific archetypes, which will extend the product-line archetypes. Now the relations between the archetypes need to be defined. One thing to reflect over is if archetypes overlap the overlap should be removed if possible.

Describing the product instantiations is the final step and its goal is to verify the suitability of the selected archetypes and the ability of the architecture to represent all variations of the product. All products at the extremes need to be described in the product instantiations, which also allows studying and addressing the conflicts between features that remain after the scoping. Some of the conflicts may have been missed, most probable are those for the least typical products, but they does not necessarily need to be addressed during the design but can be dealt with during the design of the product containing the conflict.

As per discussion made above, we can say that the method of architectural design followed for the naval product line of CelsiusTech [4] is Architectural Based Design. With reference to the literature, other successful case studies that follow the other method of architectural design i.e., Functionality based architectural design can be studied. Thus, in this section, we provided a bird's eye view of product line architecture design issues and discussed how CelsiusTech evolved with its product line architecture based on the above mentioned methods of architectural design. The next section deals with the UML profile for product lines explained with a simple example.

## 5. A UML Profile for Product Lines

The Unified Modeling Language (UML) [6] is a standard for object-oriented analysis and designing. It defines a set of notations to describe different aspects of a system. Use cases, sequence diagrams, class diagrams, component diagrams and statecharts are examples of these notations. UML is a large and regrettably complex language. Still, there are many requests to explicitly represent additional features that cannot be described comfortably with UML in its current version. Therefore, the UML provides mechanisms, in particular *stereotypes* and *tagged values* that allow extensions. These extensions may be defined and grouped in so called *profiles*.

Software Product Line engineering aims at improving productivity and decrease realization times by gathering the analysis, design and implementation activities of a family of systems. Variabilities are characteristics that may vary from a product to another. The main challenge in the context of software Product Lines (PL) approach is to model and implement these variabilities. Even if the product line approach is a new paradigm, managing variability in software systems is not a new problem and some design and programming techniques allows handling variability.

UML includes some techniques such as inheritance, cardinality range, and class template that allow the description of variability in single product i.e. variability is specified in the product models and resolved at run time. Furthermore, it is interesting to use UML to specify and to model not only one product but a set of products. In this case the UML models should be considered as reference models from which product models can be derived and created. This variability corresponds to the product line variability. We use UML extension mechanisms to specify this product line variability in UML class diagrams and sequence diagrams. A set of *stereotypes*, *tagged values* and structural *constraints* are defined and gathered in a UML profile for PL.

## 6. Summary and Conclusion

This paper presented an architecture based development paradigm known as software product lines. UML profile is been discussed to support variations in the products. The product line approach is steadily climbing in popularity as more organizations see true order-of-magnitude improvements in cost, schedule, and quality from using it. Like all new technologies, however, this one holds some surprises for the unaware. Architecturally, the key is identifying and managing commonalities and variations, but non technical issues must be addresses as well, including how the organization adopts the model, structures itself, and maintains its external interfaces.

To say, by adopting product line approach, CelsiusTech achieved great trademark in building complex software-intense systems. They observed a great shrinkage in schedules, code reusability by using core assets during this product line development. Also they expanded their business area from naval uses to air force by taking the advantage of the architecture that was originally developed for naval uses. Many successful case studies can be discussed if we look into the literature.

Along with the benefits, there are also some difficulties involved in developing the product line in any organization. Product line practice, like any new technology, needs careful thought given to its adoption, and a company's history, situation, and culture must be taken into account. The factors that contribute to product line failure can be-

- Lack of champion in a position of sufficient control and visibility
- Failure of management to provide sustained and unwavering support
- Failure to clearly identify business goals for adopting the product line approach
- Failure to adequately train staff in the approach and failure to explain or justify the change adequately.

However, by adopting a product line involved education and training on the part of management and technicians, CelsiusTech succeeded in its approach.